\documentclass[journal]{IEEEtran}
\usepackage{amsmath,amsfonts}
\usepackage{algorithmic}
\usepackage{array}
\usepackage{textcomp}
\usepackage{stfloats}
\usepackage{url}
\usepackage{verbatim}
\usepackage{graphicx}
\def\BibTeX{{\rm B\kern-.05em{\sc i\kern-.025em b}\kern-.08em
    T\kern-.1667em\lower.7ex\hbox{E}\kern-.125emX}}
\usepackage{balance}
\usepackage{makecell}
\usepackage[square, comma, sort&compress, numbers]{natbib}
\usepackage{mathrsfs}
\usepackage{booktabs}
\usepackage{xcolor}
\newcommand{\stress}[1]{{\color{black} #1}}
\usepackage[linesnumbered,ruled]{algorithm2e}
\usepackage{subfigure}

\ifCLASSINFOpdf
\else
\fi

\begin{document}
\title{FoV and Efficiency Optimization for Resonant Beam SLIPT with Telescope Integration}

\author{

Qingwei Jiang,
Shun Han,
Mingliang Xiong,~\IEEEmembership{Member,~IEEE},
Mengyuan Xu,
Zeqian Guo,
Wen Fang,~\IEEEmembership{Member,~IEEE},
and Qingwen Liu,~\IEEEmembership{Senior Member,~IEEE}

\thanks{Copyright (c) 20xx IEEE. Personal use of this material is permitted. However, permission to use this material for any other purposes must be obtained from the IEEE by sending a request to pubs-permissions@ieee.org.}
\thanks{This work was supported in part by the National Key Research and Development Program of China under Grant 2022YFA1004700; in part by the National Natural Science Foundation of China under Grants 62305019, 62671431, 62603507, and 62301308; and in part by the China Postdoctoral Science Foundation under Grant 2025M781657. \textit{(Corresponding author: Mingliang Xiong.)}}
\thanks{Q.~Jiang, S.~Han, M.~Xiong, Z. Guo, and Q.~Liu are with the School of Computer Science and Technology, Tongji University, Shanghai 201804, China (e-mail: jiangqw@tongji.edu.cn, hanshun@tongji.edu.cn, mlx@tongji.edu.cn, guozeqian@tongji.edu.cn and qliu@tongji.edu.cn).}
\thanks{M.~Xu is with the College of Electronic and Information Engineering and the Shanghai Research Institute for Intelligent Autonomous Systems, Tongji University, Shanghai, China, and also with the Shanghai Artificial Intelligence Laboratory, Shanghai, China (e-mail: xumy@tongji.edu.cn).}
\thanks{W.~Fang is with the College of Electronic and Information Engineering, Tongji University, Shanghai 201804, China (e-mail: wen.fang@tongji.edu.cn).}

}


\maketitle

\begin{abstract}

Meeting the large bandwidth demands of wireless communication for mobile Internet of Things (IoT) devices while enhancing their endurance is a significant challenge.
Simultaneous lightwave information and power transfer (SLIPT) technology offers the potential to realize wireless charging and signal transfer, making it suitable for supporting autonomous vehicles and drones.
The resonant beam system (RBS) leverages the self-aligning property of a spatially distributed laser resonator (SSLR), allowing energy transmission from the transmitter to the receiver without mechanical alignment. 
However, the existing resonant beam SLIPT system exhibits a limited field of view (FoV) and transmission efficiency, facing challenges in practical applications. 
In this paper, we propose a resonant beam SLIPT system incorporating an internal telescope and establish a performance optimization model for the system. By co-optimizing the parameters of the RBS and the internal telescope configuration, enhancement in both power transfer and FoV performance is achieved.
The results indicate that the optimized FoV is increased by $17\%$, reaching $\pm26.8^\circ$, while its average end-to-end efficiency is improved by $145\%$, achieving $5.4\%$.

\end{abstract}


\section{Introduction}
\IEEEPARstart{S}imultaneous lightwave information and power transfer (SLIPT) has attracted wide attention in extending the battery lifetime and providing wireless communication connection of energy-constrained terminal equipment (e.g., smartphones, electric vehicles, robots, drones, etc.)~\cite{r2,r3,r4}.
Additionally, it is regarded as a candidate solution for facilitating the transition to the 6G era and advancing sustainable Internet of Things (IoT) technologies~\cite{9770377}.
SLIPT utilizes light-emitting diodes (LEDs) or lasers to enable wireless connection with mobile devices, offering the benefits of a broader communication bandwidth while providing a continuous power supply to these devices~\cite{r5,r6}.

Integrating energy harvesting modules into the optical wireless communication (OWC) systems facilitates the implementation of SLIPT technology.
The most promising OWC technologies include visible light communication (VLC), optical camera communication (OCC), free-space optics (FSO), and resonant beam systems (RBS)~\cite{jahid_contemporary_2022,r22,periyathambiExploringExperimentalPossibilities2024,r7,r8}.
Among these, VLC technology is the most mature, typically utilizing divergent light sources such as LEDs and LDs to achieve wireless communication~\cite{aboagyeRISassistedVisibleLight2023}.
As early as 2019, the IEEE Task Group released a revised version of the VLC standard, IEEE 802.15.7-2018, which specifies the modulation schemes, forward error correction, line coding, as well as data rates for short-range optical channels in local and metropolitan area networks~\cite{IEEEStandardLocal2019}.
Combined with energy harvesting technology, these systems can wirelessly charge receiving devices~\cite{obeed_optimizing_2019, CARVALHO2014343}. 
However, their energy transmission efficiency is constrained by receiver size and beam energy density, typically remaining at the milliwatt level~\cite{5778623,7524747}.
OCC technology offers compatibility with augmented reality and virtual reality platforms, enabling offline video, audio, and text sharing between users~\cite{hossan_human_2019}, yet it suffers from low data rates and challenges in efficient energy harvesting~\cite{7890427}.
FSO technology uses collimated laser beams to support high-speed wireless communication and high-power energy transfer~\cite{limUnderwaterSLIPTPrototype2024}. 
However, its reliance on acquisition, tracking, and pointing (ATP) mechanisms to mitigate alignment errors increases system complexity and cost, rendering it less economically viable for IoT applications~\cite{nguyen_dynamic_nodate}. 
As a promising 6G technology, resonant beam systems (RBS) can effectively address the limitations of conventional OWC technologies and enable simultaneous high-speed data transmission and high-power delivery. 

\begin{figure}[!t]
\centering
\includegraphics[scale=1.12]{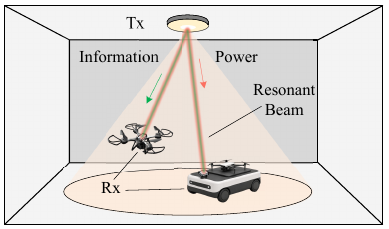}
\caption{{\stress{The example of application scenario for resonant beam SLIPT system.}}}
\label{scenario}
\end{figure}


As an intra-cavity laser, resonant beam has a small beam divergence angle, allowing for the transmission of energy up to several tens of watts with a high energy density~\cite{r11}.
Through intra-cavity frequency doubling, RBS can achieve energy and data separation, facilitating high-speed wireless communication~\cite{xiong_performance_2022}. 
Moreover, compared to other OWC systems, the RBS system benefits from its spatially distributed laser resonator (SSLR) design, which enables mobile self-alignment and offers inherent safety for human exposure~\cite{r19,fangSafetyEvaluationSelfprotection2022}. 
These advantages make RBS particularly suitable for application scenarios that demand both high mobility and stringent safety requirements~\cite{zhang_6g_2019}. 
As illustrated in Fig.~\ref{scenario}, the deployment of miniaturized RBS receivers on unmanned aerial vehicles (UAVs) or drones enables the transmitter to simultaneously deliver power and establish network connectivity to mobile devices within its operational coverage. This architecture substantially extends the endurance and enhances the communication bandwidth of energy-constrained mobile platforms.
Beyond aerial applications, RBS technology also demonstrates significant potential in indoor IoT ecosystems~\cite{r11}. In smart home or industrial IoT settings, RBS can function as a unified power and data backbone for distributed sensor networks. Low-power sensors and embedded devices can operate perpetually without battery replacement, while maintaining reliable optical communication links. Furthermore, the inherent safety characteristics of resonant beam transmission make it particularly suitable for human-centric environments~\cite{fangSafetyEvaluationSelfprotection2022}.

Despite these advantages, several challenges continue to hinder the widespread deployment of RBS.
One key limitation is the restricted field of view (FoV), which directly affects the operational range within which the receiver can move. 
Based on the original RBS structural design, the system's FoV achieved in experimental settings can reach up to $\pm 5^\circ$, which is significantly narrower compared to that of VLC technologies~\cite{r11}. 
The study in~\cite{hanCoverageExpansionResonant2025} enhanced the FoV of the RBS transmitter to $28^\circ$ by incorporating an inverted telescope. However, this modification resulted in a reduction of transmission efficiency, decreasing the transmission range to 1 meter.
Another critical issue is the system's end-to-end transmission efficiency.
In~\cite{r21}, an optimized asymmetric cavity design was proposed, which improved the system's end-to-end efficiency through structural optimization. 
Additionally, the study in~\cite{baiResonantBeamSWIPT2023} introduced a telescope internal modulator, which significantly enhanced transmission efficiency and extended the working distance to approximately 100 meters. Moreover, the application of an aspheric lens for spherical aberration correction in~\cite{sheng_continuous-wave_2021} further increased laser transmission efficiency to 36\%.
\stress{However, such improvements often come at the cost of a reduced FoV. These results reveal an intrinsic trade-off between FoV enhancement and end-to-end efficiency in conventional resonant beam systems. Therefore, a critical challenge remains in quantitatively analyzing the coupling relationship between FoV and efficiency, and further realizing their joint optimization within the RBS framework, which is the core motivation of this work.}


This paper presents a resonant beam SLIPT system incorporating an intra-cavity telescope. Through optimization of the SSLR layout, a more efficient resonator configuration is realized.
Building upon this architecture, a comprehensive performance analysis model is established for the resonant beam SLIPT system. The coupling relationship between the system's FoV and transmission efficiency is quantitatively characterized via analytical modeling and numerical evaluation.
Furthermore, a multi-objective optimization framework is developed for the system. By determining the Pareto frontier of achievable performance regions, the structural parameters of the SSLR are optimized, resulting in simultaneous enhancement of FoV, transmission efficiency, and communication data rate.
The contributions of our work are as follows:
\begin{enumerate}
    \item {\stress{We propose a novel SSLR structure by introducing an internal telescope, which spatially separates the transmitter and receiver to form an open resonant cavity. This design achieves inherent self-alignment without ATP, supports high-efficiency SLIPT, and enables flexible joint optimization of transmission efficiency and FoV.}}
    \item We develop a comprehensive performance analysis model for the resonant beam SLIPT system with an integrated intra-cavity telescope. By analytically evaluating the system’s FoV and transmission efficiency, the coupling relationship between these two performance metrics is quantitatively characterized. 
    \item We formulate and solve the optimization of the dedicated system to maximize the power delivery and FoV performance while ensuring stable communication.
    The results demonstrate that the optimized system achieves a maximum FoV of $\pm26.8^\circ$ with an average end-to-end transmission efficiency of $5.4\%$.
    Compared to the unoptimized baseline, the FoV and transmission efficiency are improved by 17\% and 145\%.
\end{enumerate}

The remainder of this paper is organized as follows. 
Section II introduces an RB-SLIPT system with an internal telescope structure. We analyze the system's energy transmission power and communication rate when the receiver is in motion. 
In Section III, we propose an optimization method based on the RB-SLIPT system and solve for the Pareto frontier of the system's achievable performance region. 
Section IV analyzes the optimized system's performance enhancements and the trade-off between power transfer and communication. 
Finally, we conclude in Section V.

\section{System Model}
\label{section2}

\subsection{RB-SLIPT Structure}

\begin{figure*}[!t]
\centering
\includegraphics[scale=1]{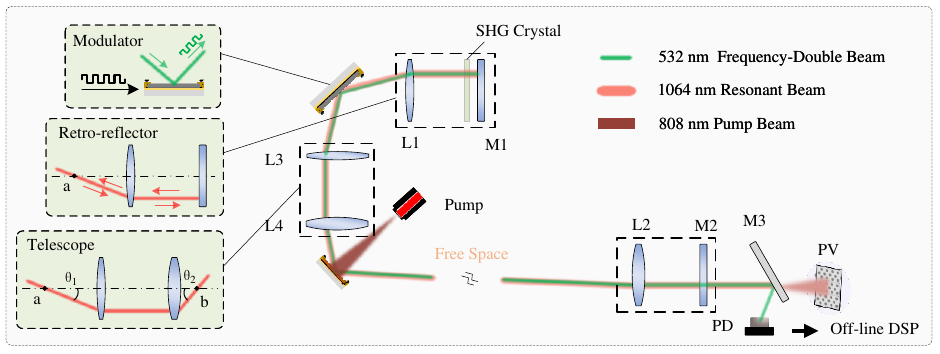}
\caption{Schematic diagram of resonant beam SLIPT system design with an internal telescope. (PD: photodiode; PV: photovoltaic)}
\label{system01}
\end{figure*}

The resonant beam communication scheme using frequency-doubled beams for simultaneous energy and data transmission was first proposed in~\cite{r13}. 
By employing frequency separation to distinguish between the power transfer and communication beams, the impact of echo interference on communication can be effectively eliminated~\cite{xiong_performance_2022}. 
In this paper, we optimize the resonant beam SLIPT system design by incorporating a telescope at the transmitter based on the design of the frequency-doubled RBS communication system. 
As shown in Fig.~\ref{system01}, the telescope is an optical module consisting of two lenses. 
The incident angle $\theta_2$ of the beam entering from the side with the shorter focal length lens will always be greater than the exit angle $\theta_1$ of the beam leaving from the other side of the telescope, and the relationship always satisfies
\begin{equation}
    \left| {\frac{{{\theta _1}}}{{{\theta _2}}}} \right| = \left| {\frac{{{f_4}}}{{{f_3}}}} \right|,
\end{equation}
where $f_3$ and $f_4$ are the focal lengths of the lenses L3 and L4 in the telescope. 
Moreover, when the telescope is placed in an appropriate position within the SSLR, the resonant beam will always pass through two points in the telescope: the anterior focal point of the front lens, denoted as point $a$, and the posterior focal point of the rear lens, denoted as point $b$.
We refer to these points as the pupil, through which the resonant beam consistently passes.
Since we introduced a telescope into the SSLR, an additional pupil appeared in the system, which can be used to place optical devices. 
Therefore, we separate the modulator and the gain medium to build a more straightforward resonant beam SLIPT system, as shown in Fig.~\ref{system01}. 
Inside the SSLR are three different wavelengths of light beams: the resonant beam for energy transmission, the frequency-doubled beam for communication, and the pump beam. 
When the pump light illuminates the gain medium, the resonant beam is generated through the stimulated emission process in the SSLR.
The resonant beam oscillates back and forth in the SSLR until the mode stabilizes. 
The stable resonant beam is then emitted through the output mirror M2, passes through the dichroic mirror M3, and is converted into electrical energy by a photovoltaic (PV) cell.
The frequency-doubled beam is generated when the resonant beam passes through a second harmonic generation (SHG) crystal. 
The frequency-doubled beam is modulated using an electro-optic modulator (EOM) and subsequently reflected by mirror M3 into a photodetector (PD), converted into electric signals.

\subsection{System stability analysis}
The resonant beam SLIPT system operates only within the stable range of the SSLR.
Therefore, it is essential to determine the stable range to ensure the designed system parameters are both rational and practical.
The transmitter of the SSLR consists of a telecentric cat's eye and a telescope.
Through ABCD matrix analysis, the ray transfer matrix of the telecentric CER can be obtained as
\begin{equation}
\label{cer}
    {{\rm{\mathbf{M}}}_{{\rm{CER}}}} = \left[ {\begin{array}{*{20}{c}}
1&0\\
{ - \frac{1}{{{f_{\rm{R}}}}}}&1
\end{array}} \right]\left[ {\begin{array}{*{20}{c}}
{ - 1}&0\\
0&{ - 1}
\end{array}} \right],
\end{equation}
where
\begin{equation}
    {f_{\rm{R}}} = \frac{{{f^2}}}{{2(l - f)}},
\end{equation}
$f$ is the lens's focal length in the telecentric CER, and $l$ is the interval between the mirror and the lens in the telecentric CER.
Since an additional telescope is included at the transmitter of the system, the ray transfer matrix of the system's transmitter can be expressed as
\begin{equation}
    {{\rm{\mathbf{M}}}_{{\rm{tx}}}} = {\rm{\mathbf{M}}}_{_{{\rm{tele}}}}^{\rm{T}}{{\rm{\mathbf{M}}}_{{\rm{CER}}}}{{\rm{\mathbf{M}}}_{{\rm{tele}}}} = \left[ {\begin{array}{*{20}{c}}
1&0\\
{ - \frac{{f_3^2}}{{{f_{{\rm{R1}}}}f_4^2}}}&1
\end{array}} \right]\left[ {\begin{array}{*{20}{c}}
{ - 1}&0\\
0&{ - 1}
\end{array}} \right],
\end{equation}

\begin{equation}
\begin{aligned}
{{\rm{\mathbf{M}}}_{{\rm{tele}}}} = &\left[ {\begin{array}{*{20}{c}}
1&{{f_3}}\\
0&1
\end{array}} \right]\left[ {\begin{array}{*{20}{c}}
1&0\\
{ - \frac{1}{{{f_3}}}}&1
\end{array}} \right]\left[ {\begin{array}{*{20}{c}}
1&{{f_3} + {f_4}}\\
0&1
\end{array}} \right]\\
&\left[ {\begin{array}{*{20}{c}}
1&0\\
{ - \frac{1}{{{f_4}}}}&1
\end{array}} \right]\left[ {\begin{array}{*{20}{c}}
1&{{f_4}}\\
0&1
\end{array}} \right]
\end{aligned},
\end{equation}

\begin{equation}
\begin{aligned}
{\rm{\mathbf{M}}}_{_{{\rm{tele}}}}^{\rm{T}} = &\left[ {\begin{array}{*{20}{c}}
1&{{f_4}}\\
0&1
\end{array}} \right]\left[ {\begin{array}{*{20}{c}}
1&0\\
{ - \frac{1}{{{f_4}}}}&1
\end{array}} \right]\left[ {\begin{array}{*{20}{c}}
1&{{f_3} + {f_4}}\\
0&1
\end{array}} \right]\\
&\left[ {\begin{array}{*{20}{c}}
1&0\\
{ - \frac{1}{{{f_3}}}}&1
\end{array}} \right]\left[ {\begin{array}{*{20}{c}}
1&{{f_3}}\\
0&1
\end{array}} \right]
\end{aligned},
\end{equation}
To ensure that the FoV of the CER and the telescope at the transmitter of the system match, it is necessary to satisfy $f_1 = f_3$. 
Furthermore, if we let $\Delta_1=l_1-f_1$, $\Delta_2=l_2-f_2$, and $M_{\rm t} = {f_3 \mathord{\left/
 {\vphantom {1 }} \right. \kern-\nulldelimiterspace}f_4}$, the ray transfer matrices of the transmitter ${\rm{\mathbf{M}}}_{\rm tx}$ and the receiver ${\rm{\mathbf{M}}}_{\rm rx}$ can be expressed as
\begin{equation}
\label{Mtx}
{{\bf{M}}_{{\rm{tx}}}} = \left[ {\begin{array}{*{20}{c}}
1&0\\
{ - {1 \mathord{\left/
 {\vphantom {1 {\frac{{f_1^2}}{{2{\Delta _1}}}}}} \right.
 \kern-\nulldelimiterspace} {\frac{{f_3^2}}{{2{\Delta _1}{M_{\rm t}^2}}}}}}&1
\end{array}} \right]\left[ {\begin{array}{*{20}{c}}
{ - 1}&0\\
0&{ - 1}
\end{array}} \right],
\end{equation}

\begin{equation}
\label{Mrx}
{{\bf{M}}_{{\rm{rx}}}} = \left[ {\begin{array}{*{20}{c}}
1&0\\
{ - {1 \mathord{\left/
 {\vphantom {1 {\frac{{f_2^2}}{{2{\Delta _2}}}}}} \right.
 \kern-\nulldelimiterspace} {\frac{{f_2^2}}{{2{\Delta _2}}}}}}&1
\end{array}} \right]\left[ {\begin{array}{*{20}{c}}
{ - 1}&0\\
0&{ - 1}
\end{array}} \right].
\end{equation}
According to Equation~(\ref{cer}), it can be found that the transmitter can be equivalently regarded as a telecentric CER with an effective focal length $f_{\rm R1}={\frac{{f_1^2}}{{2{\Delta _1}{M_{\rm t}}}}}$ and the receiver is a telecentric CER with an effective focal length $f_{\rm R2}={\frac{{f_2^2}}{{2{\Delta _2}}}}$. 
Therefore, the entire system can be considered as an SSLR composed of two asymmetric telecentric CERs.
Using the effective focal lengths $f_{\rm R1}$ and $f_{\rm R2}$, the single-pass transmission matrix of the SSLR can be further simplified from Equation~(8) in \cite{hanCoverageExpansionResonant2025} as
\begin{equation}
\begin{aligned}
    {{\bf{M}}_{\rm{s}}} =& \left[ {\begin{array}{*{20}{c}}
{ - \frac{{{\Delta _2}}}{{{f_2}}}}&{{f_2}}\\
{ - \frac{1}{{{f_2}}}}&0
\end{array}} \right]\left[ {\begin{array}{*{20}{c}}
1&d\\
0&1
\end{array}} \right]\left[ {\begin{array}{*{20}{c}}
0&{ - {f_4}}\\
{\frac{1}{{{f_4}}}}&{\frac{{{\Delta _1}}}{{{f_4}}}}
\end{array}} \right]\\
=&\left[ {\begin{array}{*{20}{c}}
{\frac{{{\Delta _2}\left( {2{f_{\rm R2}} - d} \right)}}{{{f_4}{f_2}}}}&{ \frac{{{\Delta _1}{\Delta _2}}}{{{f_4}{f_2}}}\left( {2{f_{\rm R2}} - d} \right) + \frac{{{\Delta _2}{f_1}}}{{{f_2}}}}\\
{- \frac{d}{{{f_4}{f_2}}}}&{\frac{{{\Delta _1}\left( {2{f_{\rm R1}} - d} \right)}}{{{f_4}{f_2}}}}
\end{array}} \right]
\end{aligned},
\end{equation}
while
\begin{equation}
\left\{\begin{array}{l}
{g_1}^* = \frac{{{\Delta _2}\left( {2{f_{\rm R2}} - d} \right)}}{{{f_4}{f_2}}}\\
{g_2}^* = \frac{{{\Delta _1}\left( {2{f_{\rm R1}} - d} \right)}}{{{f_4}{f_2}}}
\end{array}\right..
\end{equation}

Based on the single-pass transmission matrix ${{\bf{M}}_{\rm{s}}}$, the stability conditions of the RBS can be obtained by solving the resonant cavity stability condition $0 < {g_1}^*{g_2}^* < 1$~\cite{a181218.01}.

\subsection{Resonant beam radius calculation}
The beam radius is an important parameter of the resonant beam, which determines the diffraction loss of the resonant beam passing through each optical module in the cavity, as well as the pumping efficiency of the pump light.
The radius of curvature of the equi-phase plane and the beam radius of the resonant beam at any position in the separated resonant cavity can be expressed by introducing the q-parameter of the Gaussian beam~\cite{a181224.01}
\begin{equation}
    \frac{1}{{q(z)}} = \frac{1}{{R(z)}} - \frac{{i\lambda }}{{\pi \omega _{00}^2(z)}},
\end{equation}
where $\lambda$ is the wavelength of the resonant beam, ${R(z)}$ and $\omega _{00}(z)$ are the radius of curvature of the equi-phase plane and beam radius of the resonant beam at the propagation distance $z$, respectively.
When $z=0$, we can obtain the q-parameter of the resonant beam at mirror M1 $q(0)$.
Since mirror M1 is a plane mirror and the radius of curvature of the intra-cavity resonant beam at the mirror is always equal to the mirror's radius of curvature, we can obtain $\frac{1}{R(0)}=0$.
Thus the q-parameter of the resonant beam at M1 $q(0)$ is given by 
\begin{equation}
    q(0) =  \frac{i\pi\omega _{00}^2(0)}{\lambda},
\end{equation}
where
\begin{equation}
    \omega _{00}^2(0) = \frac{{\lambda {L^*}}}{\pi }\sqrt {\frac{{g{}_2^*}}{{g{}_1^*(1 - g_1^*g_2^*)}}},     
\end{equation}
$\omega_{00}(0)$ is the resonant beam radius at mirror M1.
According to the ABCD law, we can obtain the q-parameter of the resonant beam at the gain medium $q(l_1+f_1)$ from $q(0)$ 
\begin{equation}
    q(l_1+f_1) = \frac{{ - \frac{{{\Delta _1}}}{{{f_1}}}}q(0)+l_1}{{ - \frac{1}{{{f_1}}}}q(0)+1}.
\end{equation}

Furthermore, we can calculate the ${\rm TEM}_{00}$ mode radius of the resonant beam at the gain medium as
\begin{equation}
    {\omega _{00}}({l_1} + {f_1}) = \sqrt { - \frac{\lambda }{{\pi {\mathop{\rm Im}\nolimits} [{1 \mathord{\left/
 {\vphantom {1 {q({l_1} + {f_1})}}} \right.
 \kern-\nulldelimiterspace} {q({l_1} + {f_1})}}]}}},
\end{equation}
where ${\rm Im}[\cdot]$ represents the imaginary part of a complex number.

\subsection{Energy transfer performance analysis}

To analyze the optical power of the resonant beam in the RBS system, we need to determine the losses within the SSLR when the receiver moves.
These losses mainly consist of two parts: 1) the diffraction efficiency $\eta_t$ of the resonant beam for a round trip within the cavity; 
2) the resonant beam's scattering and absorption losses $\eta_s$ inside the gain medium, lenses, and mirrors.
Once the internal cavity losses are determined, the resonant beam intensity within the cavity can be calculated using the circulating power model~\cite{a181221.01}.

\begin{figure}[!t]
\centering
\includegraphics[scale=0.9]{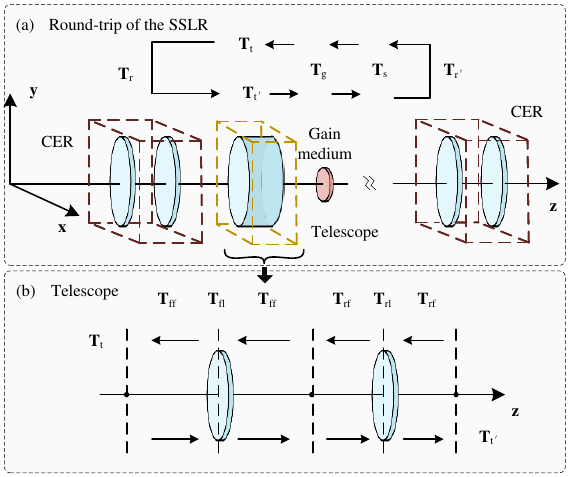}
\caption{Diagram of (a) SSLR round-trip and (b) telescope transmission matrix.}
\label{roundtrip}
\end{figure}

\subsubsection{Diffraction loss calculation}
Firstly, we solve the transmission losses within the resonant cavity by establishing a mathematical model for the proposed system. 
Figure~\ref{roundtrip} illustrates the round-trip diffraction-based transmission model of the SSLR. As shown in Fig.~\ref{roundtrip}(a), a single round-trip in the SSLR can be described as the process where an optical field starting from any plane within the cavity propagates through optical components such as lenses and mirrors and returns to the original plane. The propagation of the optical field through each optical module can be represented by a transmission matrix ${\rm T}$. A more detailed telescope model is provided in Fig.~\ref{roundtrip}(b), which consists of two lenses with different focal lengths.
As illustrated in Fig.~\ref{roundtrip}(a), we define the gain medium as the $Z=0$ plane and $u(x,y,z_0)$ as the transverse mode distribution of the resonant beam at the $Z=z_0$ plane. 
We adopt the angular spectrum theory of beam transmission~\cite{fourier} and the Fox-Li iterative method~\cite{1969Huygens} to solve for the stable transverse mode distribution of the resonant beam.
The round-trip transmission process of the resonant beam within the cavity can be represented by a self-consistent equation
\begin{equation}
\label{self}
    \begin{array}{l}
\gamma {\rm{\textbf{u} = \textbf{T}\textbf{u}}}\\
{\rm{\textbf{u}  =  }}{ {\textbf{u}(x,y,0)}^{\rm{T}}},{\rm{\textbf{T} = }}{{\rm{\textbf{T}}}_{\rm{g}}}{{\rm{\textbf{T}}}_{\rm{s}}}{{\rm{\textbf{T}}}_{\rm{r'}}}{{\rm{\textbf{T}}}_{{\rm{s}}}}{{\rm{\textbf{T}}}_{\rm{g}}}{{\rm{\textbf{T}}}_{{\rm{t'}}}}{{\rm{\textbf{T}}}_{\rm{r}}}{{\rm{\textbf{T}}}_{\rm{t}}}
\end{array},
\end{equation}
where $\gamma$ is the eigenvalue indicating the field change with one round-trip transmission.
By solving the self-consistent equation, we can obtain the value of $\gamma$ and subsequently calculate the intra-cavity diffraction efficiency ${\eta _{\rm{t}}} = {\left| \gamma  \right|^2}$. 
${\rm \textbf{T}_g}$, ${\rm \textbf{T}_r}$,  ${\rm \textbf{T}_{r'}}$,${\rm \textbf{T}_s}$, and ${\rm \textbf{T}_{s'}}$ are the transmission matrices of the optical field transfer through gain medium, CER at transmitter, CER at receiver, and free space from the receiver pupil to transmitter pupil, respectively, which are given in~\cite{r11}.
${\rm \textbf{T}_t}$ denotes the transmission matrix of the internal telescope within the cavity, and ${\rm \textbf{T}_{t'}}$ denotes the transmission matrix of the reverse telescope as in Fig.~\ref{roundtrip}(b), which can be expressed as 
\begin{equation}
    {{\bf{T}}_{\rm{t}}}{\rm{ = }}{{\bf{T}}_{{\rm{ff}}}}{{\bf{T}}_{{\rm{fl}}}}{{\bf{T}}_{{\rm{ff}}}}{{\bf{T}}_{{\rm{rf}}}}{{\bf{T}}_{{\rm{rl}}}}{{\bf{T}}_{{\rm{rf}}}},
\end{equation}
\begin{equation}
    {{\bf{T}}_{{\rm{t'}}}}{\rm{ = }}{{\bf{T}}_{{\rm{rf}}}}{{\bf{T}}_{{\rm{rl}}}}{{\bf{T}}_{{\rm{rf}}}}{{\bf{T}}_{{\rm{ff}}}}{{\bf{T}}_{{\rm{fl}}}}{{\bf{T}}_{{\rm{ff}}}},
\end{equation}
where ${{\bf{T}}_{{\rm{fl}}}}$ and ${{\bf{T}}_{{\rm{rl}}}}$ are the transition matrices for field transfer through lens L3 and lens L4, respectively.
${{\bf{T}}_{{\rm{ff}}}}$ and ${{\bf{T}}_{{\rm{rf}}}}$ are the transition matrices for field transfer through the free space with distances $f_3$ and $f_4$, respectively.
The corresponding optical field transmission operations can be expressed as
\begin{equation}
    \begin{array}{l}
{{\bf{T}}_{{\rm{rl}}}}{\bf{u}} = {\left[ {u(x,y){L_1}(x,y)} \right]^{\rm{T}}}\\
{{\bf{T}}_{{\rm{fl}}}}{\bf{u}} = {\left[ {u(x,y){L_2}(x,y)} \right]^{\rm{T}}}\\
{{\bf{T}}_{{\rm{rf}}}}{\bf{u}} = {\left[ {{{\cal F}^{ - 1}}\left\{ {{\cal F}\{ u(x,y)\} H({v_x},{v_y};{f_3})} \right\}} \right]^{\rm{T}}}\\
{{\bf{T}}_{{\rm{ff}}}}{\bf{u}} = {\left[ {{{\cal F}^{ - 1}}\left\{ {{\cal F}\{ u(x,y)\} H({v_x},{v_y};{f_4})} \right\}} \right]^{\rm{T}}}
\end{array},
\end{equation}
where ${H({v_x},{v_y};{f_3})}$ and ${H({v_x},{v_y};{f_4})}$ are the free-space propagation kernels of the field mode.
$L_1(x, y)$ and $L_2(x, y)$ represent the phase changes after the field passes through lens L3 and lens L4, respectively.
Through the Fox-Li iterative method, we can obtain the transverse mode distribution of the resonant beam in the gain medium in the stable state and the eigenvalue $\gamma$ of Equation~(\ref{self}). 
The diffraction loss ${\eta _{\rm{t}}}$ can be obtained according to $\gamma$.

\subsubsection{Scattering and absorption losses analysis}
Since the reflective or transmissive coatings on the surfaces of the optical components exhibit different reflectivity and transmittance with varying incident angles, each optical component's reflectivity and transmittance vary with different incident angles, especially at the gain medium. 
In our model, we consider the loss factor $\eta_s$ as the product of $\eta_c$ and $\eta_g$, where $\eta_c$ represents a constant that does not change with the incident angle, and $\eta_g$ represents a loss factor that changes with the incident angle. 
The loss factor $\eta_g$ is given by
\begin{equation}
    {\eta _g}(\theta ) = {\eta _{\rm{m}}}\left( {{\theta _{set}} + \frac{\theta }{{{M_{\rm{t}}}}}} \right),
\end{equation}
where ${\theta _{\rm set}}=45^\circ$ is the angle between the gain medium and optical axis of the transmitter within the SSLR and ${\eta _{\rm{m}}}\left( {\cdot} \right)$ is the transmission efficiency of the resonant beam at the gain medium. 
Since the front and rear surfaces of the gain medium are coated with anti-reflective and high-reflective films at $45^\circ$, the gain medium has the highest transmission efficiency at an incident angle of $45^\circ$.
\subsubsection{Transmission power calculation}
Based on the circulating power model, the intra-cavity resonant beam power $P_{{\rm{rb}}}$ can be depicted as~\cite{a181218.01}
\begin{equation}
\label{P_rb}
    {P_{{\rm{rb}}}(\theta)} = \frac{{{\eta_o}{\eta _c {\eta _g}(\theta )}\sqrt {{\eta _t}} \left( {{P_{{\rm{in}}}} - {P_{{\rm{th}}}}} \right)}}{{\left[ {1 - R_{\rm out}\sqrt {{\eta _t}}  + \sqrt {R_{\rm out}{\eta _t}} \left( {\frac{1}{{{\eta _c {\eta _g}(\theta )}\sqrt {{\eta _t}} }} - {\eta _s}} \right)} \right]}},
\end{equation}
where
\begin{equation}
    {P_{{\rm{th}}}} = \frac{{{a_g}{I_s}\left| {\ln \sqrt {R_{\rm out}\eta _c^2{\eta _g^2}(\theta){\eta _t}} } \right|}}{{{\eta _e}}},
\end{equation}
$I_{\rm{s}}$ is the saturation intensity, $R_{\rm out}$ is the reflectivity of the output mirror M2, ${\eta_o}$ is the overlapping efficiency, $P_{{\rm{th}}}$ is the power threshold of the input pump power, ${a_g}$ is the radius of the stimulated gain medium, and ${\eta _e}$ is the pump efficiency.

The intra-cavity resonant beam is then released by mirror M2 partially and collected by the PV, and converted into electrical energy to charge the device battery.
The PV panel can be equivalent to a current source parallel to a diode.
When a load with resistance $R_{\rm{L}}$ is connected to the PV panel, the output electrical power ${P_{\rm{pt}}}$ can be obtained as:
\begin{equation}
\label{P}
    \left\{ \begin{array}{l}
        {I_{\rm{p}} = \eta_{\rm pv}(1-R_{\rm out})P_{\rm{rb}}}\\
        {P_{\rm{pt}}} = {I_{\rm{out}}}^2{R_{\rm{L}}}\\
        {I_{\rm{out}}} = {I_{\rm{p}}} - {I_0}\left[ {{e^{\frac{{{I_{\rm{out}}}q({R_{\rm{L}}} + {R_{\rm{s}}})}}{{kTnn_{\rm{s}}}}}} - 1} \right] - \frac{{{I_{\rm{out}}}({R_{\rm{L}}} + {R_{\rm{s}}})}}{{{R_{\rm{sh}}}}}
        \end{array} \right.,
\end{equation}
where $\eta_{\rm pv}$ is the responsivity of PV, $q$ is the electron charge, $T$ is the temperature of the environment, $k$ is the Boltzmann constant, $I_0$ is the reverse saturation current, $n$ is the diode ideality factor, and $n_{\rm{s}}$ is the number of cells inside the PV panel. $R_{\rm{s}}$ and $R_{\rm{sh}}$ are the series resistance and the shunt resistance in the equivalent PV panel model, respectively.
Utilizing a maximum power point tracking device, the load resistance $R_{\rm{L}}$ of the PV can be adjusted to track the maximum power point of the PV, thus maximizing the electrical power $P_{\rm{pt}}$ outputting at the receiver.

\subsection{Communication performance analysis}

When the intra-cavity resonant beam passes through the SHG crystal, a portion of the resonant beam is converted into the frequency-doubled beam and is subsequently modulated after passing through the modulator.
We use quadrature amplitude modulation (QAM) and optical orthogonal frequency division multiplexing (O-OFDM) techniques to achieve signal modulation. 
We implement signal modulation and demodulation using intensity modulation and direct detection (IM/DD) methods. 
By applying Hermitian symmetry to the QAM symbol blocks, the OFDM signal generated by the IFFT operation is converted into a real-valued signal for use in the IM/DD system. 
At this point, the spectral efficiency $C$ of the system can be expressed as~\cite{tsonev_towards_2015}
\begin{equation}
    C = \frac{{\sum\limits_{k = 1}^{\frac{1}{2}{N_{{\rm{fft}}}}} {{{\log }_2}({M_k})} }}{{{N_{{\rm{fft}}}} + {N_{{\rm{cp}}}}}},
\end{equation}
where $N_{{\rm{fft}}}$ is the FFT size, ${M_k}$ is the constellation size modulated on the subcarrier with index $k$, $N_{{\rm{cp}}}$ is the OFDM cyclic prefix length.
The communication rate $W$ of the system can be calculated as
\begin{equation}
    W = {W_{\rm{b}}}C,
\end{equation}
where $W_{\rm b}$ is the bandwidth of the EOM.
To ensure that the communication capacity can meet the requirements for transmitting the modulated signal, the condition to be satisfied can be expressed as~\cite{lapidoth_capacity_2008}
\begin{equation}
\label{C}
\left\{ \begin{array}{l}
C < \frac{1}{2}{\log _2}\left[ {1 + \frac{{{{({\eta _{{\rm{pv}}}}{P_{{\rm{it}}}})}^2}}}{{2\pi e{R_{\rm{b}}}\left( {2q{I_{{\rm{bg}}}} + \frac{{4kT}}{{{R_{\rm{L}}}}} + 2q{\eta _{{\rm{pv}}}}{P_{{\rm{it}}}}} \right)}}} \right]\\
{P_{{\rm{it}}}} = \frac{{8\pi d_{{\rm{eff}}}^2l_{\rm{s}}^2}}{{{\varepsilon _0}c{\lambda ^2}n_0^3\omega _{\rm{s}}^2}}{P_{{\rm{rb}}}}
\end{array} \right.
\end{equation}
where $P_{{\rm{it}}}$ is the power of the frequency-doubled beam, $e$ is Euler's number, $k$ is the Boltzmann constant, $q$ is the elementary charge, $I_{{\rm{bg}}}$ is the background irradiance, $T$ is the temperature, and $R_{L}$ is the load resistance of the receiver circuit, $d_{\rm eff}$ is the SHG crystal’s effective nonlinear coefficient, $l_{\rm s}$ is the thickness of the SHG crystal, $n_0$ is the refractive index of the SHG crystal, $\varepsilon _0$ is the vacuum permittivity, $c$ is the speed of light, $\omega _{\rm s}$ is the resonant beam radius on the SHG crystal. 

\section{SYSTEM OPTIMIZATION}

In this section, we define and discuss the RB-SLIPT system's achievable performance region of the communication capacity, transmission power, and FoV performance. 
In addition, an optimization method for the resonant beam transmission system is proposed, which can be used to calculate the Pareto front of the system's achievable performance range. \stress{ We exclude frequency-doubled optical power from the set of optimization parameters. Since this output originates from nonlinear energy conversion of the fundamental beam, boosting its power inevitably compromises the overall end-to-end energy transfer efficiency; meanwhile, the obtainable communication bandwidth already satisfies prevailing application demands and can be readily modified by varying the thickness of the SHG crystal, which renders this quantity better suited as a design constraint instead of an optimization target.}

\subsection{Achievable Performance Region}

The performance of the RB-SLIPT system encompasses energy transfer power $P_{\rm pt}$, communication rate $W$, and FoV $\theta_{\rm{max}}$. 
To characterize the power transfer and communication performance of the system under mobile conditions, we define the system's FoV $\theta_{\rm{max}}$, average power transfer efficiency ${\eta _{{\rm{avg}}}}$, and average communication rate ${R_{{\rm{avg}}}}$, which can be expressed as
\begin{equation}
    {\theta _{{\rm{max}}}} = \max \left\{ {{\theta _{{\rm{tx}}}}|{P_{{\rm{pt}}}}({\theta _{{\rm{tx}}}}) > 0} \right\}
\end{equation}
\begin{equation}
    {\eta _{{\rm{avg}}}} = \int_{ - \frac{{{\theta _{{\rm{max}}}}}}{2}}^{\frac{{{\theta _{{\rm{max}}}}}}{2}} {\frac{{{P_{{\rm{pt}}}}({\theta _{{\rm{tx}}}})}}{{{\theta _{{\rm{max}}}}{P_{{\rm{in}}}}}}} d{\theta _{{\rm{tx}}}}
\end{equation}
\begin{equation}
    {R_{{\rm{avg}}}} = \int_{ - \frac{{{\theta _{{\rm{max}}}}}}{2}}^{\frac{{{\theta _{{\rm{max}}}}}}{2}} {\frac{{W({\theta _{{\rm{tx}}}})}}{{{\theta _{{\rm{max}}}}}}} d{\theta _{{\rm{tx}}}}
\end{equation}
By calculating the maximum off-axis angle $\theta_{\rm{tx}}$ at which the SSLR can maintain stable oscillation, $\theta_{\max }$ can be determined. 
Within the FoV, the resonant beam is proportionally divided into communication and power transfer beams by SHG crystals. 
The power of the 532-nm beam used for communication and the 1064-nm beam used for energy transfer affect the system's communication and energy transfer performance. 
These two performance metrics are interdependent, such that an increase in one often results in a decrease in the other. 
By calculating ${\eta _{{\rm{avg}}}}$ and ${R_{{\rm{avg}}}}$ within the system's FoV, the overall performance of the system can be effectively characterized.

\begin{figure}[!t]
\centering
\includegraphics[scale=1.3]{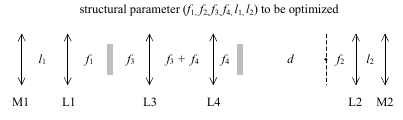}
\caption{Structural parameters to be optimized for the SSLR.}
\label{optPara}
\end{figure}

According to Equation~(\ref{P_rb}), the intra-cavity resonant beam varies with changes in parameters $\eta_{\rm o}, \eta_{\rm c}, \eta_{\rm g}, \eta_{\rm t}, a_{\rm g}$, and $R_{\rm out}$, thereby affecting $P_{\rm it}$, $P_{\rm pt}$ at different angles. 
This variation also impacts the threshold pump power, influencing $\theta_{\rm{max}}$.
Among these parameters, $\eta_{\rm o}, \eta_{\rm c}, \eta_{\rm g}$ are intrinsic to the optical components. Thus, we can enhance system performance by adjusting the values of $\eta_{\rm t}, a_{\rm g}$ and $R_{\rm out}$.
SSLR can be simplified to the model shown in Fig.~\ref{optPara}, where the main parameters to be optimized are $({{f_1},{f_2},{f_3},{f_4},{l_1},{l_2}})$.
By altering the system structural parameter $({{f_1},{f_2},{f_3},{f_4},{l_1},{l_2}})$ of the SSLR, we can modify the modes of the resonant beam in the cavity, which in turn changes $\eta_{\rm t}$.
In addition, $M_{\rm t}$ changes the loss $\eta_{\rm g}$ under different incident angles of the resonant beam, which also affects the system performance.
From the analysis of equivalent resonant cavities, it has been observed that an SSLR can be effectively modeled as a double-concave stable cavity composed of two concave mirrors. 
The radii of curvature $f_{{\rm{R}}1}$ and $f_{{\rm{R}}2}$ of the equivalent concave mirrors are determined by the parameters $(f_1,f_3,f_4,l_1)$ and $(f_2,l_2)$, respectively. 
Consequently, the system parameter $\mathbf{v_1}$ can be simplified as
\begin{equation}
    \mathbf{v_1}=(f_{{\rm{R}}1}, f_{{\rm{R}}2}, M_{\rm t}, a_{\rm g}, R,l_{\rm s}).
\end{equation}

\begin{figure}[!t]
\centering
\includegraphics[scale=0.8]{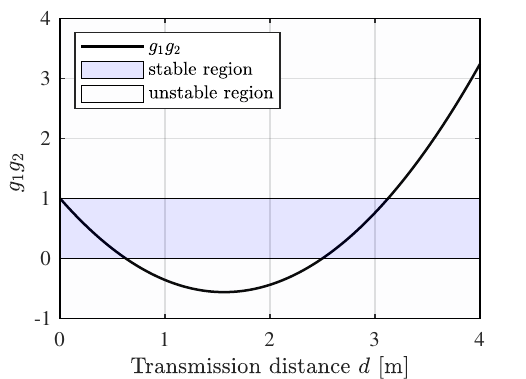}
\caption{The stable range of the SSLR.}
\label{Curve_stability_vs_dis}
\end{figure}

As system parameters vary, the stability region of the SSLR also changes. Fig.~\ref{Curve_stability_vs_dis} illustrates the evolution of system stability as the transmission distance gradually increases from 0 to $d_{\max}$. The black curve represents the trajectory of the $g_1g_2$ parameter with increasing transmission distance. The system remains stable when the curve lies within the stable region.
Otherwise, stability is lost. 
If the system fails to maintain stability within the transmission distance range of 0 to $d_{\max}$, it is considered unable to meet the design requirements.
To ensure the system's proper functioning, it is necessary to eliminate this issue, ensuring that the stability region is continuous and completely covers the maximum operating distance.
Therefore, the set of feasible RB-SLIPT system parameters $\mathbf{v_1}$ selection strategies $S$ considering stability constraints can be formulated as

\begin{equation}
    S = \{f_{{\rm{R}}1}, f_{{\rm{R}}2}, M_{\rm t}, a_{\rm g}, R, l_{\rm s}\}:\forall d \in [ 0,d_{\max} ],0 < g_1g_2 < 1,
\end{equation}
where $d_{\max}$ is the maximum operating distance of the RB-SLIPT system.

\textit{Definition 1 (Achievable performance region):}
The achievable performance region $\mathcal{R} \subseteq {\mathbb{R}_+^{3}}$ is defined as

\begin{equation}
\mathcal{R} = \left\{ {\eta _{{\rm{avg}}},{\theta _{\max }},R_{{\rm{avg}}}} \right\}:\forall \{f_{{\rm{R}}1}, f_{{\rm{R}}2}, M_{\rm t}, a_{\rm g}, R\} \in S    
\end{equation}

For the RB-SLIPT system, points within set $\mathcal{R}$ that offer superior performance lie on the Pareto front of the set $\mathcal{R}$, as follows. 
Therefore, we should determine the Pareto front of the set $\mathcal{R}$ to optimize the system.

\textit{Definition 2 (Pareto front):} A point $\textbf{x}$ is called a Pareto-optimal solution of a set $\rm{P}$, if $\textbf{x} \in {P}$ while $\textbf{x}' \in \mathbb{R}_+^L$ satisfying $\textbf{x}' > \textbf{x}$ with an element-wise inequality not in set $P$.
The set of all Pareto optimal solutions is called the Pareto front of the set $P$.

\subsection{RB-SLIPT system Optimization}
\label{Optimization}

Based on the analysis above, the system optimization problem can be transformed into maximizing the average transmission efficiency $\eta _{{\rm{avg}}}$, average communication rate $R_{{\rm{avg}}}$, and ${\theta _{\max }}$, which can be expressed as
\begin{equation}
\label{equ:1}
    \begin{aligned}
    {\rm{P}}1: &\quad \max_{\mathbf{v_1} \in \mathbb{R}_+^6} \left\{ {{\theta _{\max }},\eta _{{\rm{avg}}}, R_{{\rm{avg}}}} \right\} \quad \\
    \text{s.t.} & \quad 0 < g_1g_2 < 1, \forall d \in [ 0,d_{\max} ], \\
    & \quad P_\text{in} = P_\text{set}.
\end{aligned}
\end{equation}
where $P_\text{set}$ is the predetermined pump light intensity.
We set $d_{\max} = 6$ because a maximum operating distance of 6~m can satisfy the needs of the majority of SLIPT applications. 
The frequency-doubled beam's optical power determines the system's communication capacity. However, once the optical power of the frequency-doubled beam reaches a certain level, the communication rate of the system becomes constrained by the modulator's bandwidth and modulation format. 
Therefore, we transform Problem (P1) into an optimization problem for ${\theta _{\max }}$ and $\eta _{{\rm{avg}}}$, subject to the constraint of meeting the communication rate requirements:
\begin{equation}
\label{equ:2}
    \begin{aligned}
    {\rm{P}}1^*: &\quad \max_{\mathbf{v_1} \in \mathbb{R}_+^6} \left\{ {{\theta _{\max }},\eta _{{\rm{avg}}}} \right\} \quad \\
    \text{s.t.} & \quad 0 < g_1g_2 < 1, \forall d \in [ 0,d_{\max} ], \\
    & \quad R_{{\rm{avg}}} \ge  R_{{\rm{lb}}}, \\
    & \quad P_\text{in} = P_\text{set}.
\end{aligned}
\end{equation}
where $R_{\rm lb}$ is the lower bound of the system's communication rate.
Due to many parameter variables in $v_1$, we decompose the problem into three independent sub-problems: 1) optimizing the loss factor $\eta_{\rm t}$ under a fixed $M_{\rm t}$; 2) optimizing $\eta_{\rm avg}$ and ${\theta _{\max }}$; 3) calculating the minimum SHG crystal size that ensures the average communication rate $R_{{\rm{avg}}}$ exceeds $R_{{\rm{lb}}}$.

\subsubsection{Optimization of loss factor $\eta_{\rm t}$}
\label{optimization}
In the SSLR, the loss factor $\eta_t$ primarily includes losses from resonant beam propagation through air and diffraction losses through optical components.
The transmission losses of resonant beams in air are relatively minor and are generally considered solely related to transmission distance. 
The majority of intra-cavity losses arise from diffraction losses through optical components. 
Using larger-radius optical lenses and mirrors can reduce diffraction losses significantly, especially when the radius is two to three times larger than the radius of the resonant beam's beam spot. 
Typically, the beam spot radius of the resonant beam is on the millimeter scale, while the radii of standard lenses and mirrors are on the centimeter scale. 
Hence, diffraction losses at the lenses and mirrors are minimal. 
However, at the gain medium, increasing the radius of the gain medium requires a larger pump light spot area to achieve pumping of the gain medium, which reduces the power density of the pump light, thereby decreasing the pumping efficiency. 
Therefore, optimizing the structure of the resonator to reduce the beam spot radius at the gain medium can decrease intra-cavity transmission losses and enhance system transmission efficiency $\eta_t$.

In a specific application scenario, the relative distance between the transmitter and the receiver is generally stable. 
Therefore, we optimize the system structure for a predetermined transmission distance $d_\text{set}$. 
By setting $M_{\rm t}$ as a fixed value $M_\text{set}$, we minimize the size of the resonant beam spot at the gain medium by optimizing the parameter $(f_\text{R1},f_\text{R2})$. 
The optimization problem for the resonant beam spot aperture can be expressed as

\begin{equation}
\label{equ:1.1}
    \begin{aligned}
    {\rm{P}}1.1: &\quad (f^*_\text{R1},f^*_\text{R2}) = \underset{(f_\text{R1},f_\text{R2})}{\arg\min}\,\, \omega_\text{g} \\
    \text{s.t.} & \quad 0 < g_1g_2 < 1, \forall d \in [0, 6], \\
    & \quad d = d_\text{set},\\
    & \quad M_{\rm t} = M_\text{set}.
\end{aligned}
\end{equation}

\subsubsection{Optimization of transmission efficiency and FoV}
With the confirmed structure of the transceiver's retro-reflector, it is essential to optimize the radius of the gain medium further $a_{\rm{g}}$, the output efficiency of the output mirror $R_{\rm{out}}$, and the telescope lens ratio $M_{\rm{t}}$ within the SSLR to enhance system performance. 
$M_{\rm t}$ affects the intra-cavity diffraction losses $\eta_{\rm t}$ of the SSLR at different radial angles; 
$a_{\rm{g}}$ influences the diffraction losses of the resonant beam on the gain medium $\Gamma_g$ and the small-signal gain coefficient ${\rm g}_0$ of the gain medium; 
$R_{\text{out}}$ directly impacts the system's end-to-end transmission efficiency $\eta_{\rm avg}$ and the threshold of the pump light intensity $P_{\rm th}$. 
We need to further adjust these three parameters to optimize $\eta _{{\rm{avg}}}$ and $\theta _{\max }$ which can be formulated as

\begin{equation}
\label{equ:1.2}
    \begin{aligned}
    {\rm{P}}1.2: &\quad {a^*_\text{g},R^*_\text{out}, M^*_{\rm t}} = \underset{a_\text{g},R_\text{out},M_{\rm t}}{\arg\max}\,\, \left\{ {\eta _{{\rm{avg}}},{\theta _{\max }}} \right\} \\
    \text{s.t.} & \quad f_\text{R1},f_\text{R2} = f^*_\text{R1},f^*_\text{R2},\\
    & \quad l_{\rm s} = l_{\rm set},
\end{aligned}
\end{equation}
where $l_{\rm set}$ is the preset thickness of the SHG crystal. 
Since the loss of the intra-cavity resonant beam caused by the SHG crystal is relatively small and does not have a significant impact on $\eta _{{\rm{avg}}}$ and $\theta _{\max }$, the system communication performance can be optimized separately from the system's energy transfer and FoV performance.

\stress{\subsubsection{Optimization Settings}
The Pareto front is generated using the grid-search settings summarized in Table~\ref{tab:grid_search}. Following Algorithm~1, the optimization is conducted in three steps: beam-waist minimization through $(\Delta_1,\Delta_2)$, FoV--efficiency optimization over $(M_{\rm t},a_{\rm g},R_{\rm out})$, and SHG crystal sizing for satisfying the communication-rate constraint.}

\stress{
The diffraction efficiency $\eta_{\rm t}$ is calculated using the Fox--Li iterative method for each $(M_{\rm t},\theta)$ pair. The iteration is terminated when
\begin{equation}
\left|\gamma_{k+1}^{2}-\gamma_k^{2}\right|<10^{-4},
\end{equation}
where $\gamma^2$ denotes the round-trip diffraction efficiency. The adopted grid produces a smooth Pareto front with an FoV resolution of approximately $\pm0.1^\circ$.}

\begin{table}[t]
\centering
\caption{\stress{Grid-search settings used in the optimization.}}
\label{tab:grid_search}
\renewcommand{\arraystretch}{1.12}
\begin{tabular}{lccc}
\toprule
Variable & Search interval & Step size & Samples \\
\midrule
$\Delta_1$ & $[10^{-5},\, f_1^2/(M_{\rm t}^2d_{\max})]$~m & $10^{-5}$~m & varies \\
$\Delta_2$ & $[10^{-5},\,f_2^2/d_{\max}]$~m & $10^{-5}$~m & varies \\
$M_{\rm t}$ & $[0.5,1.7]$ & 0.1 & 13 \\
$a_{\rm g}$ & $[0.1,4]$~mm & 0.5~mm & 8 \\
$R_{\rm out}$ & $[0.5,0.99]$ & 0.05 & 10 \\
$\theta$ & $[0^\circ,30^\circ]$ & $0.2^\circ$ & 151 \\
$l_{\rm s}$ & $[0.1,10]$~mm & -- & -- \\
\bottomrule
\end{tabular}
\end{table}

\subsubsection{SHG crystal size calculation}
By solving Problem (P1.1) and Problem (P1.2) separately, we reduced the intra-cavity transmission loss by minimizing the resonant beam spot radius. Further, we selected suitable $a_\text{g}, R_\text{out}, M_{\rm t}$ to maximize $\eta _{{\rm{avg}}}$ and $\theta _{\max }$. 
Based on the optimized resonator parameters, we then determined the minimum SHG crystal size required to ensure that the system's communication rate met the minimum communication requirements, which can be expressed as
\begin{equation}
\label{equ:2.3}
    \begin{aligned}
    {\rm{P}}1.3: &\quad {l^*_{\rm s}} = \underset{l_{\rm s}}{\arg\min}\,\, \left\{ {\left| {R_{\rm avg} - R_{\rm lb}} \right|} \right\} \\
    \text{s.t.} & \quad f_\text{R1},f_\text{R2} = f^*_\text{R1},f^*_\text{R2},\\
    & \quad a_\text{g},R_\text{out},M_{\rm t} = a^*_\text{g},R^*_\text{out}, M^*_{\rm t}.\\
\end{aligned}
\end{equation}

By decomposing the optimization problem (P$1^*$) into three sub-problems, the dimensionality of the independent variable space is effectively reduced. 
Consequently, the sub-problems can be addressed individually through grid search, ultimately yielding the solution to the problem (P$1^*$). 
The specific algorithm for solving this problem can be given by Algorithm 1.
${\textbf{\rm{M}}}_{\rm{P}}$ obtained through algorithmic solution represents the Pareto front of the achievable performance region $S$, while ${\textbf{\rm{M}}}_{\rm{v_1}}$ consists of the collection of system parameters \textbf{v} that correspond to system performance in ${\textbf{\rm{M}}}_{\rm{P}}$.

\begin{figure*}[!t]	
    \centering
	\subfigure[$f_{\rm R1}$ under different $\Delta_1$ and $f_1$] 
	{
		\begin{minipage}[t]{0.3\textwidth}
			\centering          
			\includegraphics[scale=0.6]{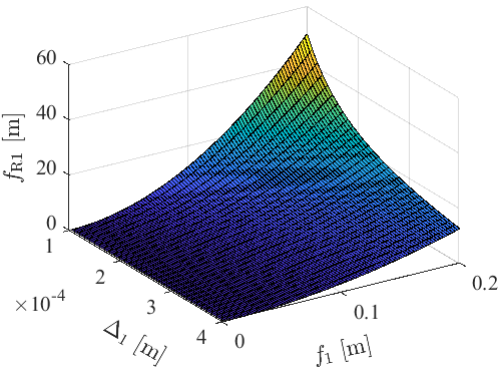}   
		\end{minipage}
	}
    \hspace{0mm} 
	\subfigure[$f_{\rm R2}$ under different $\Delta_2$ and $f_2$] 
	{
		\begin{minipage}[t]{0.3\textwidth}
			\centering      
			\includegraphics[scale=0.6]{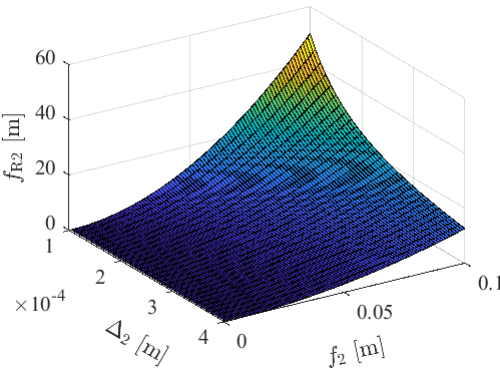}   
		\end{minipage}
	}
    \hspace{0mm} 
	\subfigure[$\omega_{\rm g}$ under different $f_{\rm R1}$ and $f_{\rm R2}$] 
	{
		\begin{minipage}[t]{0.3\textwidth}
			\centering      
			\includegraphics[scale=0.6]{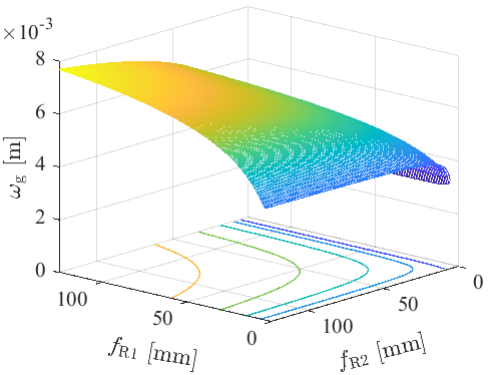}   
		\end{minipage}
	}
	\caption{Effect of ($\Delta_1$, $\Delta_2$) and focal length ($f_1$, $f_2$) on resonant beam spot aperture at gain medium $\omega_{\rm g}$.}
	\label{Curve_f_R1_VS_f_R2}  
\end{figure*}

\begin{figure}[!t]
\centering
\includegraphics[scale=1]{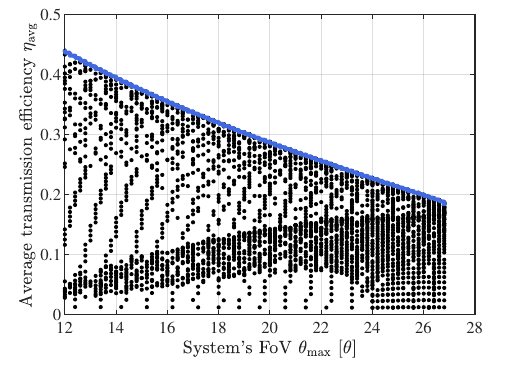}
\caption{Pareto front of the system's achievable performance region.}
\label{Curve_n_and_R_VS_FoV}
\end{figure}

\begin{algorithm}
\label{t1}
\caption{Grid Search to Optimize RBS}
\SetKwInput{KwData}{Input}
\KwData{$d_\text{set}$, $\Delta_{\max}$, $M_{\max}$, $a_{\max}$}
Initialize $\textbf{M}_1, \textbf{M}_\text{res}, {\textbf{\rm{M}}}_{\rm{v}}, {\textbf{\rm{M}}}_{\rm{p}},{\theta _{\max }}, {\omega_{\min}}$\;
\tcp{Solve Problem (P1.1)}
\For{$\Delta_1 = 0$ {\rm{to}} $\Delta_{\max}$}{
    \For{$\Delta_2 = 0$ {\rm{to}} $\Delta_{\max}$}{
        Calculate $f_\text{R1},f_\text{R2}$ according to (6)\&(7)\;
        \If{$\omega_\text{g}(f_\text{R1},f_\text{R2})<{\omega_{\min}}$}{${\omega_{\min}} = \omega_\text{g}(f_\text{R1},f_\text{R2})$\;
        $f_\text{res1},f_\text{res2} = f_\text{R1},f_\text{R2};$
        }  
    }
}
\tcp{Solve Problem (P1.2)}
\For{$M_{\rm t} = 0$ {\rm{to}} $M_{\max}$}
{
Calculate $\eta_\text{t}$ at different radial angles\;
$\textbf{M}_1$.append$((f_\text{R1},f_\text{R2},\eta_\text{t},M_{\rm t}))$\;
}
\For{itr = $1$ {\rm{to}}{
    {\rm{length}}}$(\textbf{M}_1)$}{
    $(f_\text{R1},f_\text{R2},\eta_\text{t},M_{\rm t}) = \textbf{M}_1[itr]$\;
    \For{$a_g = 0$ {\rm{to}} $a_{\max}$}{
        \For{$R_{\rm{out}} = 0$ {\rm{to}} $1$}{
            Calculate $\eta_\text{avg},\theta$ according to (26)\&(27)\;
            $\textbf{M}_\text{res}$.append$((f_\text{R1},f_\text{R2},\eta_\text{t},M_{\rm t}, a_{\rm g},\eta_\text{avg},\theta))$\;     
        }
    }
}
\tcp{Find Pareto front of the set $\textbf{M}_\text{res}$}
\tcp{Solve Problem (P1.3)}
Sort $\textbf{M}_\text{res}$ in descending order by the first element $\eta_\text{avg}$\;
\For{itr = $1$ {\rm{to}}{
    {\rm{length}}}$(\textbf{\rm{M}}_{\rm{res}})$}{
    $f_\text{R1},f_\text{R2},\eta_\text{t},M_{\rm t}, \eta_\text{avg},\theta = \textbf{M}_\text{res}[itr]$\;
    \If{${\theta}>{\theta_{\max}}$}{
        Calculate $l_{\rm s}$ according to (\ref{C})\;
        $\textbf{M}_\text{P}$.append$((\eta_\text{avg}, \theta, R_{\rm avg}))$\;
        $\textbf{M}_\text{v}$.append$((f_\text{R1},f_\text{R2},a_{\rm g},R,M_{\rm t}, l_{\rm s}))$\;
        ${\theta_{\max}} = \theta$;
    }
}
\KwResult{${\textbf{\rm{M}}}_{\rm{p}}$,${\textbf{\rm{M}}}_{\rm{v}}$}
\end{algorithm}

Based on Algorithm 1, the optimization problem P1.1 is initially solved, which involves optimizing the minimum spot radius of the resonant beam at the gain medium under stable conditions by adjusting the equivalent focal length $(f_\text{R1}, f_\text{R2})$. 
The pattern of the equivalent focal length $f_\text{R1}$ and the parameters $\Delta_1$ and $f_1$ is illustrated in Fig.~\ref{Curve_f_R1_VS_f_R2}(a), while the relationship between the equivalent focal length $f_\text{R1}$ and the parameters $\Delta_2$ and $f_2$ is depicted in Fig.~\ref{Curve_f_R1_VS_f_R2}(b).
Since both parameters $\Delta_1$ and $f_1$ influence $f_\text{R1}$, we maintain $f_1 = 25$~mm and vary $\Delta_1$ to change the value of $f_\text{R1}$.
Similarly, we set $f_1 = 25$~mm and control the variation in $f_\text{R2}$ by changing $\Delta_2$.
Figure~\ref{Curve_f_R1_VS_f_R2}(c) depicts the relationship between the resonant beam spot radius at the gain medium and the equivalent focal length $(f_\text{R1}, f_\text{R2})$. 
We can obtain the minimum spot radius and the corresponding equivalent focal length through grid search, which is
\begin{equation}
    \left\{ \begin{array}{l}
{\omega _{\rm{g}}} = {\rm{1}}{\rm{.9~mm}},\\
{f^*_{{\rm{R}}1}} = {\rm{3}}{\rm{.9063~m}},\\
{f^*_{{\rm{R}}2}} = {\rm{3}}{\rm{.0488~m}},
\end{array} \right.
\end{equation}

By substituting $({f^*_{{\rm{R}}1}},{f^*_{{\rm{R}}2}})$ into the problem (1.2) and further solving the feasible solution via grid search for the Pareto frontier, the final optimization results are shown in Fig.~\ref{Curve_n_and_R_VS_FoV}. 
The area covered by the black dots represents the range of all feasible solutions calculated, and the blue lines represent the Pareto frontier of the feasible solutions. 
In the Pareto frontier set, $\theta _{\max }$ can reach up to $\pm26.8^\circ$, and $\eta _{{\rm{avg}}}$ can reach $13.2\%$. 
Depending on different application scenarios, appropriate system parameters can be selected to design an RB-SLIPT system so that the performance can meet corresponding requirements.

\section{Numerical Results and Discussions}
 
\subsection{Parameter Settings}

The parameters of the RB-SLIPT system are listed in Table~\ref{tab:my_label}.
We use Nd:YVO$_4$ crystal as the gain medium and KTP crystal to realize SHG.
A 100~W semiconductor laser with a wavelength of 808~nm is used as a pump source to generate a 1064-nm energy transfer beam and a 532-nm communication carrier.
At the receiver, we use InGaAsP and GaAs as the materials for PV and PD, respectively, due to their excellent responsivity at 1064~nm and 532~nm.
The diameter of the intra-cavity lens and mirrors is 25.4~mm. 
To match the FoV of the transceiver retro-reflectors, we set the focal length of lenses L2 and L4 to 25.4~mm.

We employ 256-QAM to modulate the frequency-doubled beam and utilize O-OFDM technology to multiplex different subcarriers, with a total of $N_{\rm fft}= 1024$ subcarriers. 
Under these conditions, the system's spectral efficiency $C = 3.98$~bit/s/Hz. 
Electro-absorption modulators can be used as the modulator of the frequency-doubled beam, whose modulation bandwidth can reach 800~MHz~\cite{quintana2017high}.
The system's communication rate can reach up to 3.18~Gbps.
\stress{We can ensure that the system's signal-to-noise ratio meets the required communication standards by designing an appropriately sized frequency-doubling crystal as detailed in Section~\ref{optimization}~\cite{xiong,KTP}.}

\begin{table}[ht]
    \centering
    \caption{Parameters in theoretical calculation}
    \label{tab:my_label}
    \begin{tabular}{ccc}
    \toprule
    \textbf{Parameter} & \textbf{Symbol} & \textbf{Value} \\
    \midrule
    Saturation intensity & $I_{\rm{s}}$   & ${\rm{1.26}\times \rm{10^7~W/m^2}}$\\
    Effective nonlinear coefficient & $d_{\rm{eff}}$ & $4.7~\rm{pm/V}$\\
    KTP crystal refractive index & $n_0$ & $\stress{1.74}$\\
    Excitation efficiency & $\eta _{\rm{e}}$ & 72\%\\
    PV's responsivity & $\eta_{\rm{pv}}$ & $0.4~\rm{A/W}$\\
    PD's responsivity & $\eta_{\rm{pd}}$ & $0.4~\rm{A/W}$\\
    Load resistor & $R_{\rm{IL}}$ & $10 ~\text{k}\Omega$\\
    Shunt resistance & $R_{\rm{sh}}$ & $53.81 \Omega$\\
    Series resistance & $R_{\rm{s}}$ & $37 ~\text{m}\Omega$\\
    \bottomrule
    \end{tabular}
\end{table}

\subsection{Performance of the optimized resonant beam SLIPT system}

\begin{figure}[!t]
\centering
\includegraphics[scale=1]{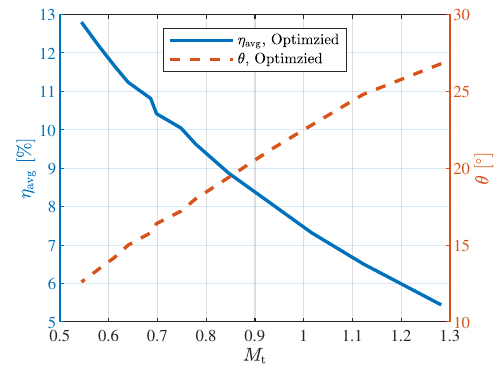}
\caption{Average transmission efficiency $\eta_{\rm avg}$ and FoV $\theta_{\rm max}$ as a function of telescope ratio $M_{\rm t}$ in the Pareto front of the system's achievable performance region.}
\label{Curve_IT_and_PT_vs_angle_0}
\end{figure}

\begin{figure}[!t]
\centering
\includegraphics[scale=1]{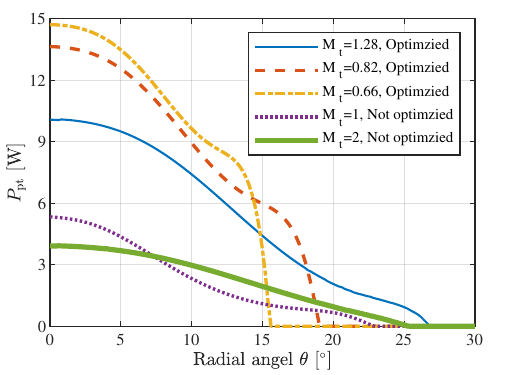}
\caption{Average transmission efficiency $\eta_{\rm avg}$ as a function of radial angle $\theta$ under different system structures.}
\label{Curve_IT_and_PT_vs_angle_1}
\end{figure}

\begin{figure}[!t]
\centering
\includegraphics[scale=1]{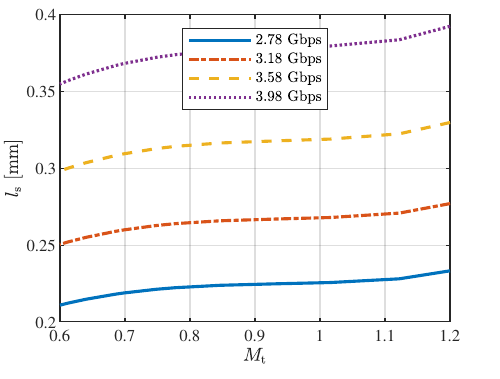}
\caption{SHG crystal thickness $l_{\rm s}$ as a function of telescope ratio $M_{\rm t}$ under different communication rate requirements.}
\label{Curve_IT_and_PT_vs_angle_4}
\end{figure}

\begin{figure}[!t]
\centering
\includegraphics[scale=1]{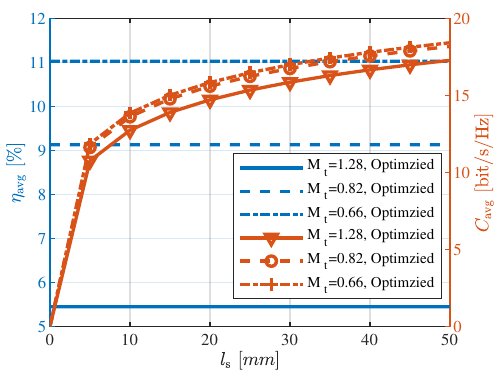}
\caption{Average power transfer efficiency $\eta_{\rm avg}$ and average communication capacity $C_{\rm avg}$ as a function of SHG crystal thickness $l_{\rm s}$ under optimized system structures.}
\label{Curve_PT_IT_VS_l_s}
\end{figure}

By applying the optimization algorithm proposed in Section \ref{Optimization}, we computed the Pareto frontier of achievable system performance at $P_{\text{in}} = 100~\text{W}$. By rearranging the Pareto front set in Fig.~\ref{Curve_n_and_R_VS_FoV}, we obtain Fig.~\ref{Curve_IT_and_PT_vs_angle_0}, which shows the FoV and average transmission efficiency $\eta_{\text{avg}}$ of the optimized system as a function of the telescope focal length ratio $M_{\text{t}}$.
Our analysis reveals that $M_{\text{t}}$ is the most critical factor influencing the trade-off between FoV and transmission efficiency in the optimized system, exhibiting a well-defined relationship. As $M_{\text{t}}$ increases, $\eta_{\text{avg}}$ continuously decreases, while the system's FoV $\theta_{\text{max}}$ progressively increases.
Notably, when $M_{\text{t}}$ exceeds 1.28, the system performance deviates from the Pareto frontier. This indicates that further increasing $M_{\text{t}}$ beyond this point no longer enhances the FoV. This phenomenon occurs because the marginal improvement in transmission efficiency at the edge of the FoV, gained by increasing $M_{\text{t}}$, is outweighed by the overall reduction in system transmission efficiency. Consequently, any further increase in $M_{\text{t}}$ leads to simultaneous degradation of both FoV and transmission efficiency.

Figure~\ref{Curve_IT_and_PT_vs_angle_1} compares the output electrical power $P_{\text{pt}}$ for different $M_{\text{t}}$ values along the Pareto frontier (1.28, 0.82, and 0.66) with unoptimized system configurations ($M_{\text{t}} = 1$ and $M_{\text{t}} = 2$). \stress{This comparison also serves as an ablation study to distinguish the effects of telescope integration and structural optimization.} The configurations $M_{\text{t}} = 0.66$, 0.82, and 1.28 on the Pareto frontier represent high transmission efficiency with small FoV, balanced performance, and large FoV with reduced efficiency, respectively. The configurations $M_{\text{t}} = 1$ and $M_{\text{t}} = 2$ correspond to systems without a telescope and with a telescope of focal length ratio 2, respectively.
Before optimization, regardless of telescope implementation, the systems suffered from relatively high transmission losses due to suboptimal structural design, resulting in low $P_{\text{pt}}$ values. \stress{The unoptimized $M_{\text{t}} = 2$ case further shows that telescope insertion alone does not guarantee substantial performance improvement.} After optimization, systems with focal length ratios of 0.82 and 0.66 show significantly improved average power transmission efficiency. Compared to the unoptimized telescope-free resonant beam SLIPT system ($M_{\text{t}} = 1$), the optimized system ($M_{\text{t}} = 0.66$) achieves a 400\% improvement in average transmission efficiency.
Furthermore, the optimization substantially enhances the system's FoV. At $M_{\text{t}} = 1.28$, the system achieves a maximum FoV of $\pm 26.8^\circ$ with an average transmission efficiency of 5.4\%. Compared to the unoptimized telescope-free system, this represents a 17\% improvement in FoV and a 145\% improvement in average transmission efficiency.
This analysis demonstrates the inherent trade-off between FoV and transmission efficiency in the system. Maximizing the FoV necessitates sacrificing transmission efficiency, while accepting a reduction in FoV enables significant gains in transmission efficiency.

Furthermore, we determined the minimum SHG crystal thickness $l_{\rm s}$ required to meet the communication rate for different telescope lens ratios $M_{\rm t}$.
Figure~\ref{Curve_IT_and_PT_vs_angle_4} illustrates the relationship between the required thickness of the SHG crystal $l_{\rm s}$ and the telescope lens ratio $M_{\rm t}$ under varying communication rate requirements. 
We employed 64~QAM, 128~QAM, 256~QAM, and 512~QAM modulation formats, achieving communication rates of 2.78~Gbps, 3.18~Gbps, 3.58~Gbps, and 3.98~Gbps, respectively.
As $M_{\rm t}$ increases, the system's average transmission efficiency decreases, resulting in a reduced $\eta_{\rm avg}$ for the same $l_{\rm s}$. 
Consequently, to maintain communication efficiency, a thicker SHG crystal is required. 
Additionally, as the communication rate increases, the required thickness of the SHG crystal correspondingly increases.
In the resonant beam SLIPT system, the communication signal-to-noise ratio of the system can be improved by increasing the optical power of the optical carrier. 
Therefore, we can control the SHG efficiency $\eta_{\rm shg}$ by adjusting the thickness of the SHG crystal $l_{\rm s}$, thereby increasing the communication rate of the system.

Figure~\ref{Curve_PT_IT_VS_l_s} illustrates the variations in power transfer capability and communication capacity under different $M_{\rm t}$ as the thickness of the SHG crystal $l_{\rm s}$ increases.
With increasing $l_{\rm s}$, the communication capacity of the system rises significantly, while the power transfer remains largely unaffected. This indicates that the power of the frequency-doubled beam used for communication constitutes only a small portion of the total resonant beam energy.
At $l_{\rm s} = 10~\text{mm}$, systems with focal length ratios $M_{\rm t}$ of 1.28, 0.82, and 0.66 achieve communication capacities of 13 bit/s/Hz, 14 bit/s/Hz, and 14.2 bit/s/Hz, respectively, thereby satisfying most practical communication requirements.

\subsection{Discussions}

\begin{table*}[ht]
    \centering
    \caption{{\stress{System performance comparison}}}
    \label{tab:my_label2}
    \begin{tabular}{cccccc}
    \toprule
    \textbf{Ref.}  & \textbf{Rx/Tx aperture [mm]} &\textbf{Working distance}& \textbf{System's FoV} &\textbf{Communication capacity}&\textbf{Transmission efficiency}\\
    \midrule
    2022~\cite{r11} & 7~/~7~mm  &2~m&$6^\circ$& --&1.7\%\\
    2023~\cite{10134572} & 2~/~2~mm &2~m&$\pm4^\circ$& 10~bit/s/Hz &$2.1$\%\\
    2025~\cite{hanCoverageExpansionResonant2025} & 25.4~/~25.4~mm &1~m&$28^\circ$& -- &0.16\%\\
    This work  & 25.4~/~25.4~mm  &6~m&$\pm 26.8^\circ$& 13~bit/s/Hz&5.4\%\\
    \bottomrule
    \end{tabular}
\end{table*}

The performance of the proposed resonant beam SLIPT system is compared with other resonant beam systems in Table~\ref{tab:my_label2}. It can be observed that through systematic optimization, the proposed system achieves nearly double the FoV while improving the transmission efficiency by an order of magnitude compared to the system in \cite{hanCoverageExpansionResonant2025}, even with an extended transmission distance of 6 meters.

Compared to other similar technologies, the RBS offers superior feasibility. The resonant beam, as an intra-cavity laser, has a small divergence angle and excellent directionality. This results in a significant increase in transmission efficiency compared to systems that use wider beams~\cite{11074846}. For systems that employ narrow beams to achieve SLIPT, the primary limiting factors are human safety, the size of the transmitter and receiver, and the complexity of the transceiver system.
RF-based SLIPT systems cannot ensure human safety, and because large-scale antenna arrays are required for beam directionality, the transceiver systems are typically bulky and unsuitable for small-scale applications such as drones or unmanned vehicles~\cite{chengAdaptiveMultiTargetMicrowave2022,10980047}. Systems that use extra-cavity lasers for SLIPT also fail to ensure human safety. Furthermore, due to the small divergence angle of the extra-cavity laser, strict alignment between the transmitter and receiver is necessary to establish a connection. This requires additional ATP mechanisms, significantly increasing the size and complexity of the transceiver~\cite{kaymakSurveyAcquisitionTracking2018}.

In the experimental setup of the resonant beam SLIPT system, the transmitter is placed in a fixed position as the base station, while the receiver is mounted on a mobile unit to receive both energy and signals~\cite{r11}. Consequently, the size of the transmitter does not need to be strictly controlled. The transmitter mainly consists of the CER, telescope, gain medium, and pump light. The aperture radius of the lenses and mirrors forming the CER and telescope is 12.7~mm. The gain medium is placed on a $5~{\rm cm} \times 5~{\rm cm}$ semiconductor cooling plate for heat dissipation. The receiver's CER also consists of lenses and mirrors with an aperture radius of 12.7~mm, housed in a cylindrical mirror tube with a radius of 12.7~mm and a length of 7~cm~\cite{hanTransmitterRotationField2023}.
Since the receiver includes only two lenses, a mirror tube, a PD, a PV cell, and a few electronic components, its size and weight are easily manageable, making it suitable for installation on small vehicles.

\stress{The main experimental challenge lies in developing a customized intra-cavity modulation module that can simultaneously support 532-nm signal modulation and 1064-nm resonant feedback. This requires compatible dual-wavelength coatings, sufficient active aperture and angular FoV, and precise coaxial alignment, which are difficult to satisfy with standard commercial modulators.
We will continue prototype validation in future work.}

\section{Conclusions}
This paper presents a resonant beam simultaneous light and power transfer (SLIPT) system incorporating an internal telescope and establishes a performance optimization model for the system. 
By incorporating a telescope at the transmitter, the modulator and gain medium can be spatially separated, thereby simplifying the system architecture.
Furthermore, we co-optimized the parameters of the resonant beam system (RBS) and internal telescope configuration to delineate the Pareto frontier of the system's achievable performance regions.
Our results demonstrate that the system's field of view (FoV) and transmission efficiency can be modulated by adjusting the focal length ratio between the front and rear lenses of the telescope. As the focal length ratio increases, the FoV progressively expands, while the transmission efficiency decreases. At a focal length ratio of 1.28, the system achieves a maximum FoV of $\pm 26.8^\circ$ with an average optical transmission efficiency of 5.4\%.
Compared with an unoptimized baseline, the resonant beam SLIPT system exhibits 17\% and 145\% improvements in FoV and transmission efficiency, respectively.

\ifCLASSOPTIONcaptionsoff
  \newpage
 \fi

\bibliographystyle{IEEEtran}
\small

\balance
\bibliography{ref}

\end{document}